\documentclass[twocolumn,showpacs,superscriptaddress,floatfix,preprintnumbers,amssymb,amsmath]{revtex4-1}

\usepackage{graphicx}
\usepackage{dcolumn}
\usepackage{bm}
\usepackage[latin1]{inputenc}
\usepackage[mathscr]{eucal}
\usepackage{epsfig}
\usepackage{rotating}
\usepackage{color}

\begin{document}

\title{Long hold times in a two-junction electron trap}

\author{A.~Kemppinen}
\email{antti.kemppinen@mikes.fi}
\affiliation{Centre for Metrology and Accreditation (MIKES), P.O.~Box 9, 02151 Espoo, Finland}
\author{S.~V.~Lotkhov}
\affiliation{Physikalisch-Technische Bundesanstalt, Bundesallee 100, 38116 Braunschweig, Germany}
\author{O.-P.~Saira}
\affiliation{Low Temperature Laboratory, Aalto University, P.O.~Box 13500, 00076 AALTO, Finland}
\author{A.~B.~Zorin}
\affiliation{Physikalisch-Technische Bundesanstalt, Bundesallee 100, 38116 Braunschweig, Germany}
\author{J.~P.~Pekola}
\affiliation{Low Temperature Laboratory, Aalto University, P.O.~Box 13500, 00076 AALTO, Finland}
\author{A.~J.~Manninen}
\affiliation{Centre for Metrology and Accreditation (MIKES), P.O.~Box 9, 02151 Espoo, Finland}

\begin{abstract}
The hold time $\tau$ of a single-electron trap is shown to increase
significantly due to suppression of environmentally assisted tunneling
events. Using two rf-tight radiation shields instead of a single one, we
demonstrate increase of $\tau$ by a factor exceeding $10^3$, up to about
10 hours, for a trap with only two superconductor (S) -- normal-metal (N)
tunnel junctions and an on-chip resistor $R$ (R-SNS structure). In the
normal state, the improved shielding made it possible to observe $\tau
\sim$~100~s, which is in reasonable agreement with the
quantum-leakage-limited level expected for the two-electron cotunneling
process.
\end{abstract}


\maketitle

Creating a quantum current standard based on frequency-driven
single-electron transport has been one of the major goals of metrology for
about 25 years~\cite{Averin1986}. In the 1990's, the efforts were focused
mainly on metallic tunnel junction devices. Relative uncertainty of
about $10^{-8}$ was reached by the seven-junction electron pump for small
currents at the level of a few pA~\cite{Keller1996}. Recent
advances~\cite{Mooij2006,Blumenthal2007,Pekola2008} have raised the hope
to increase the current by factor 10--100 towards a more practical level,
renewing the interest to the accuracy issues.

$Photon$ (aka $environmentally$) $assisted$ $tunneling$
(PAT)~\cite{Kautz2000,Pekola2010} has been found to limit the transport
accuracy at the level far above theoretical
predictions~\cite{Kautz1999,Averin2008}. Recently, it was
shown that PAT rate $\Gamma$ in the superconductor~(S) -- normal metal~(N)
hybrid turnstile~\cite{Pekola2008} can be reduced down to time-resolved
single events by using on-chip noise filtering elements: a high-ohmic
resistor (R-SNS turnstile~\cite{Lotkhov2009,Lotkhov2011}) or a capacitively
coupled ground plane~\cite{Pekola2010,Saira2010,Saira2011}. The electron
trapping properties of the pumps are important for electron-counting
capacitance standards~\cite{Keller1999}, and also a way to study
the leakage mechanisms. In the preceding experiment with the R-SNS
turnstile-based electron trapping circuit, Ref.~\cite{Lotkhov2011},
a reasonably low value of $\Gamma = \tau^{-1} \sim$~(20~s)$^{-1}$ was
achieved in a strong Coulomb blockade. Here, $\tau$ is
the average hold time of an electron in the trap.

In this Letter, we report on the very same R-SNS trap samples tested in a
cryogenic setup with a different type of noise filtering: double rf
shielding of the sample stage. We demonstrate more than 1000-fold
improvement of $\tau$ up to about 10~h. This result is comparable with the
hold times previously reported for much more complex arrays with 7 or more tunnel
junctions~\cite{Dresselhaus1994,Krupenin1997} or with 4 junctions and a
high-ohmic resistor connected in series ~\cite{Lotkhov1999}. With the
superconductivity suppressed by a magnetic field, tunneling rates down
to the level expected for $two$-$electron$
$cotunneling$~\cite{Averin19891990,Odintsov1992,Golubev1992} were achieved.

The equivalent circuit of the trapping device is shown in Fig.~1(a). The
left junction of an Al--AuPd--Al SNS turnstile is connected to an on-chip
Cr resistor with $R\sim $~100~k$\Omega$, and the right junction is
terminated by a trapping island whose charge state $n_2$ is monitored by a
capacitively coupled SNS single-electron transistor (SET) electrometer
operating in the voltage-bias regime. For the details of sample
fabrication see Ref.~\cite{Lotkhov2011}.

\begin{figure}[tb]
    \begin{center}
    \includegraphics[width=.5\textwidth]{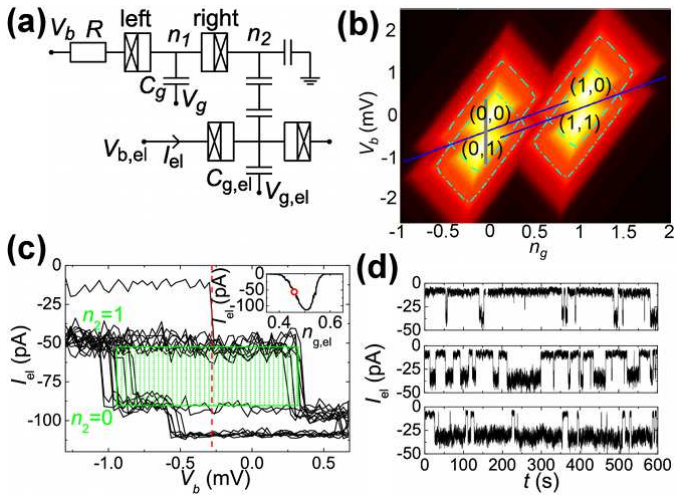}
    \end{center}
   \caption{\label{fig:basics} (Color online) (a) Circuit diagram of the trapping device. The number
   of electrons on the normal-metal (AuPd) middle island, $n_1$, and on the superconducting (Al) trap island, $n_2$, can be
   controlled by the voltages $V_{\rm b}$ and $V_{\rm g}$.
   (b) Simulated contour plot of $\tau$ at 120~mK for four charge configurations $(n_1,n_2)$ as a function
   of ($n_{\rm g}, V_{\rm b}$), $n_{\rm g} \equiv C_{\rm g} V_{\rm g}/e$. Bright and dark areas denote long and short values of $\tau$, respectively.
   In the bistable areas, the largest values of $\tau$ are plotted.
   Inclined rectangles around each stable charge state correspond to $\tau = 1$~s escape thresholds.
   Two inclined solid lines show
	the middle points of the hysteresis loops for the states $(0,0)$--$(0,1)$ and $(1,0)$--$(1,1)$.
   (c) Repeated $V_{\rm b}$ sweeps around the widest hysteresis loop of about 1.4~mV, emphasized by the rectangle 
   corresponding to the wide grey vertical line in (b).
   The red dashed line in the middle denotes the symmetric case, with equal values of $\tau$ for both charge states involved.
   Inset: Electrometer signal $I_\mathrm{el}$ as a function of gate charge $n_{\rm {g,el}}=C_{\rm {g,el}}V_{\rm {g,el}}/e$.
   The circle denotes a typical operating point.
   (d) Random switchings of the trap between two charge states. The trap is tuned to exhibit experimentally convenient hold times
   by setting the width of the hysteresis loop to about 100~$\mu$V. The central panel shows the symmetric case with $V_{\rm b}$ set in the middle of a hysteresis loop ($\Delta V_{\rm b}=0$), while
   the upper and lower traces are for $\Delta V_{\rm b}=\mp 7$~$\mu$V which  favor one of the states.
   }
\end{figure}

Two nominally similar samples, S$_{\rm A}$ and S$_{\rm B}$, were measured
in two cryostats, C$_{\rm A}$ and C$_{\rm B}$, respectively. Both sample
stages were equipped with two nested rf radiation shields. In setup
C$_{\rm A}$, the signal lines were filtered by the combination of
Thermocoax\texttrademark and powder filters, and in C$_{\rm B}$ only by
Thermocoax\texttrademark. The radiation shields of C$_{\rm A}$
were of higher quality, and unless otherwise stated, the reported results
are from S$_{\rm A}$/C$_{\rm A}$. 
During the preparation of this report, we have applied similar nested rf shielding in studies of
sub-gap behaviour of
SNS transistors~\cite{Saira2011}, and we also became aware of the work of Barends et al.~\cite{Barends2011} in which similar techniques were used to lower loss and decoherence of
superconducting quantum circuits.

At low temperatures, $T <$~100~mK, tunneling in the zero-biased trap is
blocked by the energy barrier formed by the quasiparticle excitation
energy $\Delta _{\rm {Al}} =$ 260~$\mu$eV $= k_{\rm B}
\times$~3.0~K and charging energies $E_{\rm {C1}} \approx e^2/\left [ 2
\times \left ( 2 C_{\rm T} + C_{\rm g}\right ) \right ] \approx k_{\rm B}
\times$~7.6~K~$>E_{\rm {C2}} \approx e^2/2C_{\rm {trap}} \approx k_{\rm B}
\times$~3.3~K (estimated for sample S$_{\rm A}$) of an electron located in
the middle and in the trapping island, respectively. Here $C_{\rm T}$ is
the tunnel capacitance assumed equal for both junctions, $C_{\rm g}$ is
the gate capacitance of the turnstile, and $C_{\rm {trap}}$ is the total
capacitance of the trapping island. The corresponding stability diagram is
shown in Fig.~1(b) with the hold times $\tau$ calculated using the
orthodox theory of single-electron tunneling~\cite{Averin1986} for four
charge states of the trap in thermal equilibrium.

Electrons can be loaded or unloaded from the trap by ramping the bias
voltage $V_{\rm b}$ and thus lowering the energy barrier for one of the
two tunneling directions. Due to the energy barrier, the bias voltage
dependence of the trapped charge $n_2$ monitored by electrometer current
$I_{\rm{el}}$ appears hysteretic, see Fig.~1(c). The maximum of the
barrier height and, thus, the width of the hysteresis loop can be
controlled by the gate voltage $V_{\rm g}$. Figure~1(c) shows several
ramps of $V_{\rm b}$ encircling the widest loop which corresponds to an
integer value of $n_{\rm g} \equiv C_{\rm g} V_{\rm g}/e$, see Fig.~1(b).
Due to finite $T$ and external excitations particularly addressed in this
work, the observed switchings between charge states include a random
component, leading to an uncertainty of the loop width and to the
two-level fluctuator behavior depicted in Fig.~1(d).

To support the conclusions of the long hold times shown below, we
studied $\tau$ as a function of the energy barrier (electrostatic energy change) 
of tunneling, $\Delta E$, which can be controlled by voltages
$V_{\rm b}$ and $V_{\rm g}$.
Two-state switching traces were analyzed statistically for
different values of $\Delta E$. Figure~2(a) shows typical
dependencies of $\tau$ on the deviation of bias voltage from the middle of
the hysteresis loop ($\Delta V_{\rm b}$) for three different values of the
electrometer current $I_{\rm{el}}$. When $\Delta V_{\rm b}=0$, $\Delta E$
is the same for the two values of $n_2$. A shift in $\Delta V_{\rm b}$
shifts proportionally both values of $\Delta E$, but in opposite directions.
The observed linear
dependency of log($\tau$) on $\Delta E$ is typical for thermal activation (TA, cf,
e.g., Ref.~\cite{Dresselhaus1994}): $\tau\propto \exp(-\Delta E/(k_{\rm
B}T^*))$, where the effective temperature $T^*$ characterizes the spectrum
of fluctuations driving the state switchings.

\begin{figure}[tb]
    \begin{center}
    \includegraphics[width=.5\textwidth]{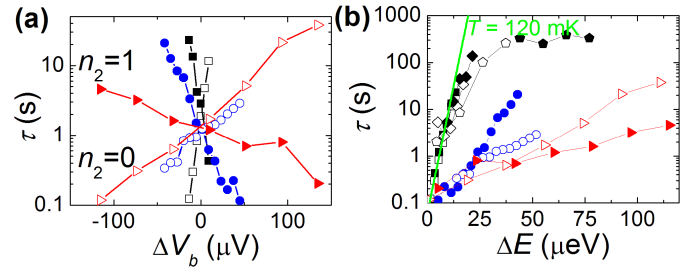}
    \end{center}
   \caption{\label{fig:X}
   (Color online)(a) Average hold time $\tau$ of two charge states as a function of bias voltage deviation $\Delta V_{\rm b}$ from the middle of the hysteresis loop
   measured at three different levels of the electrometer current: $I_{\rm{el}}<100$~pA (black squares), $I_{\rm{el}}\approx 300$~pA (blue circles), and $I_{\rm{el}}\approx 750$~pA (red triangles). Open and filled symbols of the same type correspond
to the two alternating charge states within the trace: $n_2=0$ and $n_2=1$, respectively [see Figs. 1(c--d)]. For each $I_{\rm {el}}$, $V_{\rm g}$ was adjusted such that $\tau \approx$~1~s at $\Delta V_{\rm b} = 0$.
   (b) Summarizing plot: hold time $\tau$ vs. energy barrier $\Delta E$. Green solid line shows the slope corresponding to thermal
   activation at $T^*=$~120~mK. Symbols are as in (a). Black diamonds and pentagons were measured at $I_{\rm{el}}<100$~pA with a wider hysteresis loop than black squares.}
\end{figure}

To study the crossover from TA to PAT, we present a plot
in Fig.~2(b) summarizing the hold time data $\tau$ versus $\Delta E$ for
different electrometer currents $I_{\rm{el}}$. For this approximative plot, we used
the following conventions: \emph{(i)} The origin of the 
$\Delta E$~=~0 axis was chosen arbitrarily as the point where $\tau\approx
0.1$~s. \emph{(ii)} The proportionality factor between $\Delta E$ and
$\Delta V_{\rm b}$ is based on the fitted stability diagram. \emph{(iii)}
For the curves with the smallest $I_{\rm{el}}$
(yielding the longest hold times), we combined the piecewise-continuous
data sets measured at different $V_{\rm g}$ and arranged them along the
$\Delta E$-axis in Fig.~2(b) to ensure their smooth alignment. This was
done because collecting the statistics for a short-living state would
take impractically long if following a large hysteresis loop:
switching back from the long-living counter state would take hours.

Although Fig.~2(b) is approximative, it allows us to make several conclusions.
For small $I_{\rm{el}}<100$~pA, a well distinguished slope of log($\tau$)
vs. $\Delta E$ corresponding to $T^*\approx$~120~mK is observed when $\tau
\le 100$~s. The typical electron temperature in the cryostat C$_{\rm A}$
is lower, about 50~mK, so even in this case there are still
contributing mechanisms beyond the pure TA process. For higher barriers
and longer hold times, the slope is much lower, corresponding to
$T^*\sim$~1~K. We presume that the shallow slopes are due to PAT
at frequency $f\sim \Delta E/h$, activated either by residual 
photons leaking inside the shields or by the back-action of the electrometer.
It generates random telegraph noise (RTN) whose power spectrum has a relatively
slow decay $S(\omega)\propto 1/f^2$ in the relevant frequency scale
$f\gg I_{\rm {el}}/e$ (for more details, see Ref.~\cite{Saira2011}).
The electrometer back-action is obviously dominant at the larger values of
$I_{\rm{el}}$, but it cannot be ruled out even in the case of low
$I_{\rm{el}}$ and high $\Delta E$. 

Our main result, very long hold time, was achieved with the widest hysteresis loop, i.e. the highest energy barrier $\Delta E
\sim \left ( \Delta_{\rm {Al}}+ E_{\rm {C1}} \right ) \sim k_{\rm B}
\times$10~K, and $I_{\rm {el}}<100$~pA. In that case, time interval between state switchings was more than 10 hours, which is too long 
to obtain statistically valid data. The example of Fig.~3(a) shows a
36-hour trace during which the state changed only once, 14 hours after the
start of the measurement. This should be compared with the maximum $\tau
\sim $~20~s for the same sample reported in Ref.~\cite{Lotkhov2011}. At
higher $I_{\rm{el}}$, see Fig.~3(b), the electrometer back-action imposes
a significant limitation for the maximum $\tau$. We also did some
measurements with sample S$_{\rm B}$ in cryostat C$_{\rm B}$. The maximum
hold times were of the order of 1~h. The difference was most likely due to
a lower quality of the radiation shielding.

\begin{figure}[tb]
    \begin{center}
    \includegraphics[width=.5\textwidth]{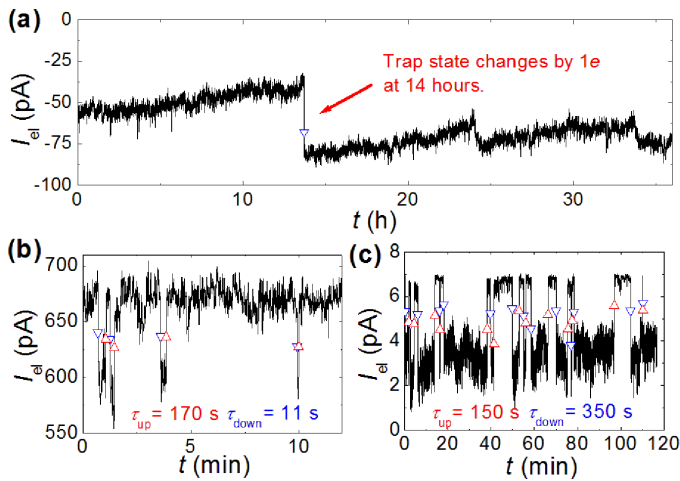}
    \end{center}
   \caption{\label{fig:maxholdtime} (a) Electrometer data trace at the maximum
   hysteresis shown in Fig.~1(c) and at low $I_{\rm{el}}$. The trap state changed only once at $t \approx 14$~h. Slow drifts in $I_{\rm{el}}$ are caused by background charge fluctuations of the electrometer.
   (b) Switching trace at the maximum hysteresis with higher $I_{\rm{el}}>$~500 pA.
   (c) Normal state switching trace for sample S$_{\rm B}$.
}
\end{figure}

The trapping times for sample S$_{\rm B}$ were studied also in the normal
state in a magnetic field in cryostat C$_{\rm B}$. A data trace near the
maximum of the hysteresis is shown in Fig.~3(c). Contrary to the NS hybrid
state, where two-electron cotunneling is suppressed by the superconducting
energy gap \cite{Pekola2008,Averin2008}, this leakage mechanism can dominate
in the normal state if PAT is suppressed strongly enough. Experimental
results are in reasonable agreement with the expected cotunneling rate
corresponding to $\tau\sim 100$~s, which we estimated following the
approximations of Ref.~\cite{Odintsov1992} for the case of dissipative
environment. For a similar structure without the resistor, the expected
value for $\tau$ is of the order of 0.1~ms. We note that the cotunneling
rates are very sensitive to the parameters of the trap, which we
have estimated rather indirectly. Hence an exact quantitative comparison
between theory and experiments would require that the parameters were
determined by measuring the $IV$ curve of the trap, for instance, with the help
of a cryogenic switch~\cite{Keller1996}. Nevertheless, our observations
indicate that a high
degree of shielding of the sample space makes it possible to reach and
study a quantum-leakage-limited behaviour of SET circuits in the normal
state.

To conclude, a 1000-fold extension of the hold times was achieved for a
two-junction R-SNS electron trap by enclosing it into a specially-designed
low-noise environment. In the normal state, a switching rate near the
quantum leakage floor set by the resistively suppressed two-electron
cotunneling was achieved. The hold times of the trap have been measured
for the same device over a wide dynamic range from 0.1 s to about $10^4$~s,
showing high sensitivity to environmental fluctuations, including the
back-action from the SET electrometer. Device application for
sensitive on-chip noise spectrometry seems feasible.

We acknowledge H. Koivula for advice in rf issues, O. Hahtela
and E. Mykk\"anen for contributions in building the measurement
system, and T. Aref for assistance in measurements. The work was partially
supported by Technology Industries of Finland Centennial Foundation and
by the Finnish Academy of Science and Letters, V\"ais\"al\"a Foundation. The
research conducted within the EURAMET joint research project REUNIAM and
EU project SCOPE has received funding from the European Community's
Seventh Framework Programme under Grant Agreements Nos. 217257 and 218783.


\end{document}